\documentclass[prd,aps,twocolumn,preprintnumbers, showpacs, nofootinbib,superscriptaddress,notitlepage]{revtex4-1}
\usepackage{amssymb,amsthm,amsmath}
\usepackage{graphicx}   % figures
\usepackage{color}      % color is used in text
\usepackage{slashed}    % Feynman slash
\usepackage{verbatim}
\usepackage[normalem]{ulem}
\usepackage{rotating}   % to rotate tables
\usepackage{multirow}   % multicolumn and multirow
\begin{document}
\title{Two-body hadronic decays of $\Xi^0_c$  in light front approach}

\author{ Hang Liu}
\email{Email:liuhang303@sjtu.edu.cn}
\affiliation{INPAC, Key Laboratory for Particle Astrophysics and Cosmology (MOE), Shanghai Key Laboratory for Particle Physics and Cosmology, School of Physics and Astronomy, Shanghai Jiao Tong University, Shanghai
200240, China}

% \author{Zhi-Peng Xing} \email{Email:zpxing@sjtu.edu.cn}
% \affiliation{Tsung-Dao Lee Institute, Shanghai Jiao Tong University, Shanghai 200240, China}

\author{Chang Yang}\email{Email:15201868391@sjtu.edu.cn}
\affiliation{INPAC, Key Laboratory for Particle Astrophysics and Cosmology (MOE), Shanghai Key Laboratory for Particle Physics and Cosmology, School of Physics and Astronomy, Shanghai Jiao Tong University, Shanghai
200240, China}
\affiliation{Tsung-Dao Lee Institute, Shanghai Jiao Tong University, Shanghai 200240, China}

\begin{abstract}
In this study, we investigate the nonleptonic decays of the charmed-baryon $\Xi^0_c$ induced by the $c\to u(d\bar d)/(s\bar s)$ transition. Utilizing the factorization assumption,  we decompose the decay amplitudes in terms of transition form factors which are then calculated within the light-front quark model.  We employ helicity amplitudes to analyze the nonleptonic decay modes of the charmed-baryons $\Xi^0_c$ and derive benchmark results for decay widths and branching fractions. Our calculations suggest that the branching fractions for some of these rare nonleptonic decays are at the order of $10^{-4}-10^{-3}$, which are likely to be detectable at experiments such as LHCb or BESIII.  The potential data accumulated in the future may help to further our understanding of the decay mechanism in the presence of charm quarks.
\end{abstract}

\maketitle

%%%%%%%%%%%%%%%%%%%%%%%
\section{Introduction}
%%%%%%%%%%%%%%%%%%%%%%%\
Weak decays of quarks can provide valuable insights into testing the standard model of particle physics and advancing our understanding of CP violation in the universe. As most quarks in nature are confined within hadrons, the study of weak decays of quarks inside a hadron also provides a unique opportunity to explore strong interactions. Over the past few decades, significant progress has been made on both the experimental and theoretical fronts, resulting in unprecedentedly precise experimental measurements and theoretical calculations of hadron decays.

Both theoretical and experimental studies have shown considerable interest in the two-body hadronic decays of charmed baryons~\cite{BESIII:2018cvs,BESIII:2018ciw,Belle:2020xku,Lyu:2021biq,He:2018joe,Wang:2017gxe,Zhao:2018mov,Geng:2019xbo,Chua:2018ikx,Hsiao:2020iwc,Du:2022ipt,Ali:1998eb}. Experimental data on these decays have been extensively collected from various sources~\cite{Belle:2018kzz,Belle:2021avh,BESIII:2020nme}, while theoretical calculations have proven challenging due to the strong QCD interaction at the charm scale. Lattice QCD is believed to ultimately offer reliable theoretical results of form factors~\cite{Zhang:2021oja,Meinel:2017ggx,Bahtiyar:2021voz,Bahtiyar:2016dom,Meinel:2021rbm,Meinel:2021mdj}. And several theoretical studies on these decays rely on modeling QCD dynamics to predict and test various models~\cite{Aliev:2021wat,Zhao:2018zcb,Zhao:2021sje,Liu:2010bh,Li:2016qai,Guo:2005qa}. Sufficient experimental data would allow for the determination of all amplitudes classified by $SU(3)$ symmetry properties, enabling systematic predictions to further test the model and gain additional insights into strong QCD interactions at the charm scale~\cite{Shi:2017dto,Huang:2021aqu,Wang:2017azm,Huang:2021jxt,Lu:2016ogy,Hsiao:2021nsc,He:2015fwa,Geng:2019bfz,Cheng:2014rfa,Sharma:1996sc,Savage:1989qr}. However, some experimental results for the specific processes are still missing. The purpose of this paper is to utilize the light front approach in computing the rare nonleptonic decays of $\Xi_c^{0}$, including those that have not yet been observed experimentally.

As the phase space for this class of decays is extremely limited, only a few channels are kinematically accessible. In this study, we investigate the $c\to u$ transition in the nonleptonic decays of heavy baryons $\Xi_c^{0}$, and consider explicitly 
\begin{itemize}
\item the spin-${1}/{2}$ to spin-${1}/{2}$ decays,
\begin{eqnarray}
&&\Xi_c^{0}\to \Sigma^0\eta,\quad\Xi_c^{0}\to \Sigma^0\eta'\notag\\
&&\Xi_c^{0}\to \Lambda\eta,\quad\Xi_c^{0}\to \Lambda\eta'\notag
\end{eqnarray}
\end{itemize}
%While the small phase space will substantially suppress these decay branching fractions, and make them very difficult to be observed in experiments, the small phase space allows for solid theoretical predictions. On the one hand, measurements of semileptonic decay widths can help to provide rather reliable constraints on the form factors since the recoil is small. With the knowledge of form factors, nonleptonic decays could give a direct exploration of theoretical  tools such as factorization. On the other hand, due to the small phase space, the  helicity suppression amplitudes in certain semi-leptonic decays are uplifted by the muon mass. These effects  are absent in semi-leptonic decays with electrons in the final state. Thus these classes of decays induced by $s\to u$  provide a new platform for studying the helicity suppression amplitudes in the experiment and offer possibilities of significant new physics (NP) contributions.

The spin of the $ds$ system in the baryons $\Xi^0_c$ and $\Xi^{0\prime}_c$ can take on values of either 0 or 1. To simplify our analysis, we use $\Xi_c^0/\Lambda/\Sigma^{0}$ to refer to the baryons with a spin-0 $ds$ system, and $\Xi_c^{\prime0}$ to denote the baryons with a spin-1 $ds$ system. In this study, we are solely focused on the decay of $\Xi_c^0$ and do not consider other decay processes of $\Xi^{0\prime}_c$.

The remaining sections of this paper are organized as follows. In Section II, we present the theoretical framework in detail. After outlining the parametrization of the form factors for spin-1/2 to spin-1/2 processes, we provide an explicit calculation of these form factors using the light-front approach. In Section III, we present our numerical results for the form factors and provide a phenomenological analysis that includes decay widths and branching ratios. Finally, we conclude our work with a brief summary in the last section.\\
 \\
%%%%%%%%%%%%%%%%%%%%%%%
\section{Theoretical  framework}
%%%%%%%%%%%%%%%%%%%%%%%

\begin{figure}[htbp!] 
\includegraphics[width=0.8\columnwidth]{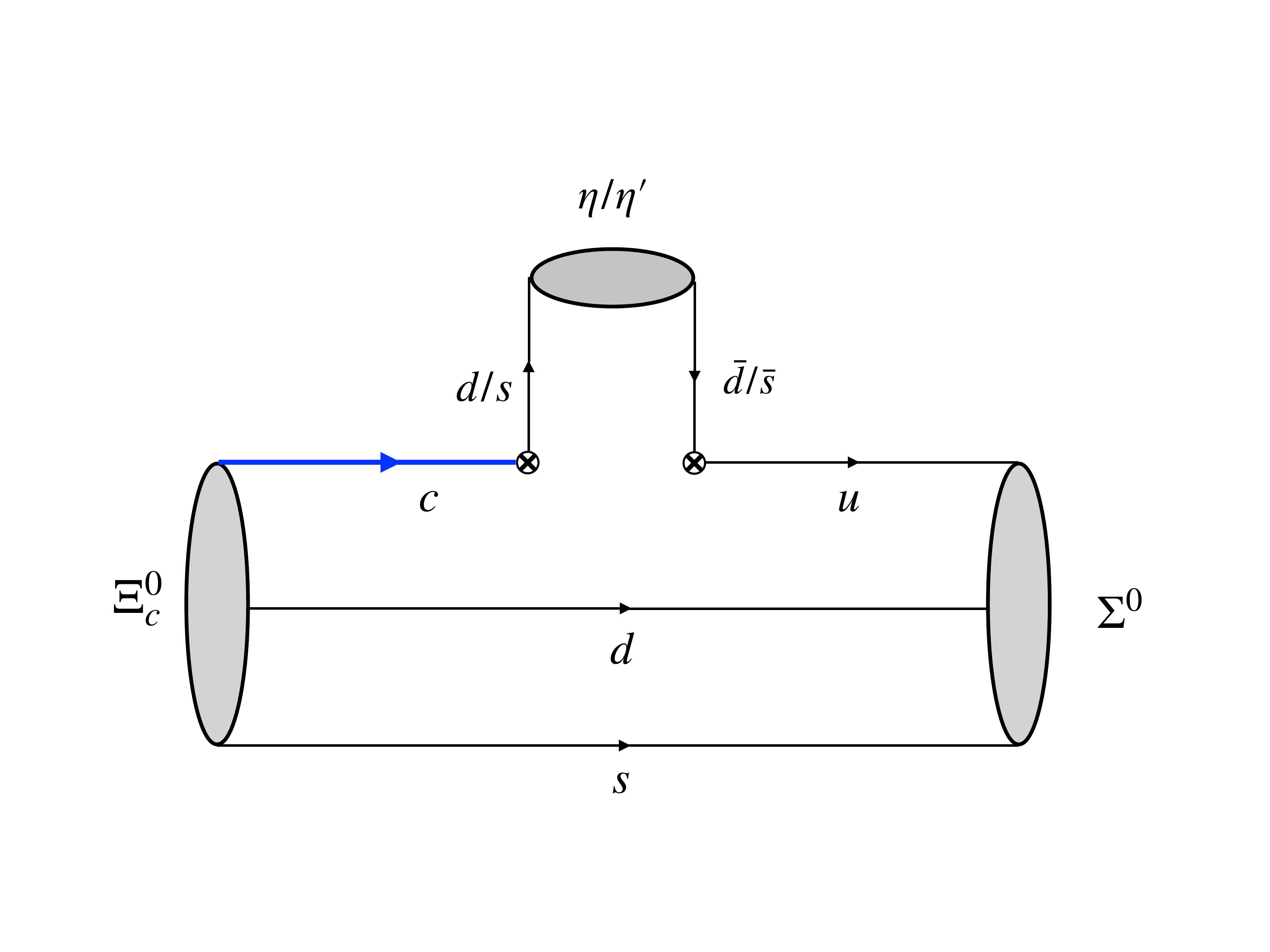}  
\caption{The Feynman diagrams for nonleptonic decays of the heavy decay baryons we investigate.}
\label{fd1}
\end{figure}

The nonleptonic decays under consideration are induced by the $c$ quark transitioning to either $u\bar{d}d$ or $u\bar{s}s$. The corresponding Feynman diagram for these processes is depicted in Fig.~\ref{fd1}.
% For the nonleptonic decay processes, we only consider the factorizable contributions, which are less reliable but can give benchmark estimate of decay branching fractions. 

%%%%%%%%%%%%%%%%%%%%%%%%%%%%%%%%%%%%%%

%%%%%%%%%%%%%%%%%%%%%%%%%%%%%%%%%%%%

The effective Hamiltonians for $c\to u \bar{d} d$ and $c\to u\bar{s}s$ are given as 
%%%%%%%%%%%%%%%%%%%%%%%
\begin{eqnarray}
&&\mathcal{H}(c\to u \bar q q)=\frac{G_F}{\sqrt{2}}V_{cq}V^*_{uq}\nonumber\\
&& \times \bigg\{ C_1 [\bar u_\alpha\gamma_\mu(1-\gamma_5)q_\beta][\bar q_\beta\gamma^\mu(1-\gamma_5)c_\alpha]\nonumber\\
&& +C_2[\bar u_\alpha\gamma_\mu(1-\gamma_5)q_\alpha][\bar q_\beta\gamma^\mu(1-\gamma_5)c_\beta]\bigg\},
\end{eqnarray}
with $q=d/s$. 
Using the Hamiltonian and Fierz identity, one can derive the nonleptonic decay amplitudes as 
\begin{eqnarray}
&&i\mathcal{M}(\Xi_c^0\to \Sigma^0\eta_d)=\frac{G_F}{\sqrt{2}}V_{cd}V^*_{ud} \nonumber\\
&&\times
a_2 \langle\eta(P_\eta)|\bar d\gamma^\mu(1-\gamma_5)d|0\rangle \nonumber\\
&&\times 
\langle \mathcal{B}_{uds}(P^\prime,S^\prime_z)|\bar u\gamma_\mu(1-\gamma_5)c|\Xi_c^0(P,S_z)\rangle,\\ 
&&i\mathcal{M}(\Xi_c^0\to \Sigma^0\eta_s)=\frac{G_F}{\sqrt{2}}V_{cs}V^*_{us} \nonumber\\
&&\times
a_2 \langle\eta(P_\eta)|\bar s\gamma^\mu(1-\gamma_5)s|0\rangle \nonumber\\
&&\times 
\langle \mathcal{B}_{uds}(P^\prime,S^\prime_z)|\bar u\gamma_\mu(1-\gamma_5)c|\Xi_c^0(P,S_z)\rangle,
\label{amp}
\end{eqnarray}
with $a_2=C_1+C_2/N_c$. The $C_i$ are Wilson coefficients at $m_c$ scale whose values can be taken from Ref.~\cite{Buras:1998raa}: $C_1=-0.636$,  $C_2=1.346$.  In the above, the $P$ and $P^\prime$ are the momentum of the initial and the final baryons $\Xi_c^0$ and $\Sigma^0$ respectively.

The hadronic contributions to these processes are described by the hadron matrix elements, which can be characterized by the form factors. In the case of spin-1/2 to spin-1/2 processes, the form factors are defined as:
\begin{widetext}
\begin{align}
&\langle\Sigma^0(P^\prime,S^\prime=\frac{1}{2},S^\prime_z)|\bar u\gamma_\mu(1-\gamma_5)c |\Xi_c^0(P,S=\frac{1}{2},S_z)\rangle\notag\\
=&\bar u(P^\prime,S^\prime_z)\bigg[\frac{M\gamma_\mu}{\bar M} f_1(q^2)+\frac{P_\mu}{\bar M}f_2(q^2)+\frac{P^\prime_\mu}{\bar M}f_3(q^2)\bigg]u(P,S_z)\notag\\
&-\bar u(P^\prime,S^\prime_z)\big[\frac{M\gamma_\mu}{\bar M} g_1(q^2)+\frac{P_\mu}{\bar M}g_2(q^2)+\frac{P^\prime_\mu}{\bar M}g_3(q^2)\bigg]\gamma_5u(P,S_z),\label{ff1/2}
\end{align}
\end{widetext}
where $\bar M=M-M^\prime$. 

%%%%%%%%%%%%%%%%%%%
\subsection{Light-front quark model}
%%%%%%%%%%%%%%%%%%%

 The quark-diquark approximation allows us to treat baryonic systems in a manner similar to mesonic systems. In weak transitions, the two spectator quarks behave as anti-quarks. This approximation has been widely employed in the study of heavy baryon decays.~\cite{Xing:2018lre,Hu:2020mxk,Zhao:2022vfr,Wang:2022ias,Liu:2022mxv}. A recent study~\cite{Ke:2019smy} explored the three-body vertex function within the light front quark model. By examining the $\Lambda_b\to\Lambda_c$ and $\Sigma_b\to\Sigma_c$ transitions, the authors found that the three-body vertex function yields results consistent with the diquark picture. This validates the use of the diquark approximation from a certain perspective.

%%%%%%%%%%%%%%%%%%%
\begin{figure}[htbp!]
  \centering
\includegraphics[width=0.5\textwidth]{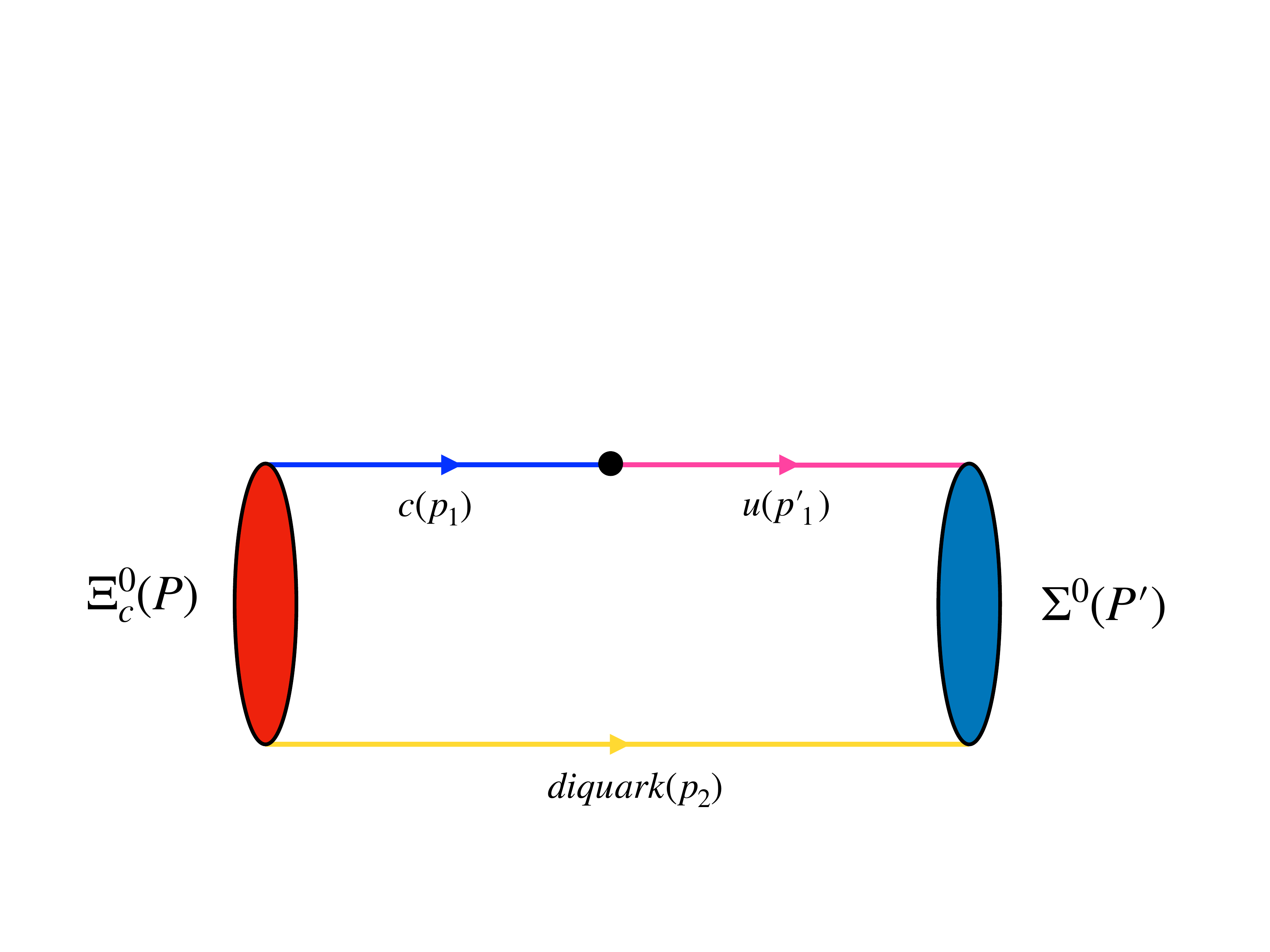}
\caption{The diquark approximation  for the baryonic transition. }
\label{diquark}
\end{figure} 
%%%%%%%%%%%%%%%%%%%%%

%%%%%%%%%%%%%%%%%%
%\subsection{Momenta space wave function}
%%%%%%%%%%%%%%%%%%%

The four-vector in the light front frame is typically represented as $v^\mu=(v^+,v^-,v_\perp)$, where $v^\pm=v^0\pm v^3$. Within the light front quark model, a baryonic state can be described by an internal quark and diquark. The hadron state can then be expanded in terms of momentum space and flavor-spin functions. For a spin-1/2 baryon state, this expansion takes the form:
\begin{eqnarray}
	&& |\Xi_c^0(P,S,S_{z})\rangle =  \int\{d^{3}p_{1}\}\{d^{3}p_{2}\}2(2\pi)^{3}\delta^{3}(\tilde{P}-\tilde{p}_{1}-\tilde{p}_{2})\nonumber \\
	&  &\;\;\;  \times\sum_{\lambda_{1},\lambda_{2}}\Psi^{SS_{z}}(\tilde{p}_{1},\tilde{p}_{2},\lambda_{1},\lambda_{2})|c(p_{1},\lambda_{1})({\rm{di}})(p_{2},\lambda_{2})\rangle,\label{eq:state_vector}
\end{eqnarray}
In the above equation, the term "$({\rm{di}})$" corresponds to the diquark shown in Fig.~\ref{diquark}, and their helicities are denoted by $\lambda_1$ and $\lambda_2$. The momenta of the baryon, quark, and diquark are represented by $P$, $p_1$, and $p_2$, respectively. The momenta $\tilde{P}$, $\tilde{p}_{1}$, and $\tilde{p}_{2}$ are three-dimensional momenta, denoted by $\tilde{p}=(p^+,p_{\perp})$. It is important to note that the on-shell momentum has only three degrees of freedom despite having four components. As a result, the minus component of the momentum is fixed by the relation $p^{-}=(m^2+p_{\perp}^2)/p^{+}$.

The wave function $\Psi$ in Eq.~\eqref{eq:state_vector} can be expressed as a combination of spin and momentum space functions. Specifically, it can be decomposed as:
\begin{eqnarray}
&&\Psi^{SS_{z}}(\tilde{p}_{1},\tilde{p}_{2},\lambda_{1},\lambda_{2})=\frac{1}{\sqrt{2(p_{1}\cdot\bar{P}+m_{1}M_{0})}} \nonumber\\
&& \;\;\; \;\;\;  \;\;\; \;\;\; \times \bar{u}(p_{1},\lambda_{1})\Gamma_{S(A)} u(\bar{P},S_{z})\phi(x,k_{\perp}).
\label{eq:momentum_wave_function_1/2}
\end{eqnarray}
As previously mentioned, the diquark can exist in either a spin-0 scalar or spin-1 axial-vector state. In the case of a scalar diquark, the interaction vertex $\Gamma$ is given by $\Gamma_S=1$. On the other hand, for a spin-1/2 baryon with an axial-vector diquark, the corresponding interaction vertex takes the form:
\begin{align}
	\Gamma_{A} & =\frac{\gamma_{5}}{\sqrt{3}}\left(\slashed\epsilon^{*}(p_{2},\lambda_{2})-\frac{M_0+m_1+m_2}{\bar{P}\cdot p_2+m_2M_0}\epsilon^{*}(p_{2},\lambda_{2})\cdot\bar{P}\right).
	\label{eq:momentum_wave_function_1/2gamma}
\end{align}

In the above equation, $m_1$ and $m_2$ represent the masses of the quark and spectator diquark, respectively. The quantity $\bar P$ is the on-shell momentum of the light quark $q$ and diquark, and satisfies the conditions $\bar{P}=p_1+p_2$ and $\bar{P}^2=M_{0}^2$. Here, $M_0$ represents the invariant mass of $\bar P$, which differs from the baryon mass $M$ due to the fact that the quark, diquark, and baryon cannot all be on their respective mass shells simultaneously. The momentum $P$ and mass $M$ of the baryon, however, must satisfy the physical mass-shell condition, $M^2=P^2$. It is important to note that the momentum $\bar P$ is not equal to $P$.

In the equation~\eqref{eq:momentum_wave_function_1/2}, $\phi$ represents a Gaussian-type function constructed as:
\begin{equation}
	\phi=4\left(\frac{\pi}{\beta^{2}}\right)^{3/4}\sqrt{\frac{e_{1}e_{2}}{x_{1}x_{2}M_{0}}}\exp\left(\frac{-\vec{k}^{2}}{2\beta^{2}}\right),\label{eq:Gauss}
\end{equation}

Here, $e_1$ and $e_2$ represent the energies of the quark $q$ and diquark in the rest frame of $\bar{P}$. The variables $x_1$ and $x_2$ correspond to the light-front momentum fractions and satisfy the conditions $0<x_2<1$ and $x_1+x_2=1$. The internal motion of the constituent quarks is described by the internal momentum $\vec k$. The internal light front  momentum and related dynamic relationships are given as:
\begin{eqnarray}
&& k_i=(k_i^-,k_i^+,k_{i\bot})=(e_i-k_{iz},e_i+k_{iz},k_{i\bot}) \nonumber\\
&& \;\;\;\; =(\frac{m_i^2+k_{i\bot}^2}{x_iM_0},x_iM_0,k_{i\bot}),\nonumber\\
&&p^+_1=x_1\bar P^+,   ~~~p^+_2=x_2 \bar P^+,  ~~~p_{1\perp}=x_1 \bar P_{\perp}+k_{1\perp}, \nonumber\\
&&  p_{2\perp}=x_2 \bar P_{\perp}+k_{2\perp},
  ~~~ k_{\perp}=-k_{1\perp}=k_{2\perp}.
\end{eqnarray}
The $\vec k$ in the Eq.~\eqref{eq:Gauss} is the internal three-momentum vector of diquark presented as $\vec k=(k_{2\bot},k_{2z})=(k_{\bot}, k_{z})$. The parameter $\beta$ appearing in Equation~\eqref{eq:Gauss} represents the momentum distribution between the constituent quarks. By using the definition of the internal three-momentum vector, we can express the invariant mass squared $M_0^2$ as a function of the variables $(x_i,k_{i\bot})$,
 \begin{eqnarray} \label{eq:Mpz}
	  M_0^2=\frac{k_{1\perp}^2+m_1^2}{x_1}+ \frac{k_{2\perp}^2+m_2^2}{x_2}.
 \end{eqnarray}
The energy $e_i$ and $k_z$ can also be expressed in terms of the internal variables $(x_i,k_{i\bot})$ as follows:
\begin{eqnarray}
&& e_i=\frac{x_iM_0}{2}+\frac{m_i^2+k_{i\perp}^2}{2x_iM_0} =\sqrt{m_i^2+k_{i\bot}^2+k_{iz}^2}, \nonumber\\
&&  k_{iz}=\frac{x_iM_0}{2}-\frac{m_i^2+k_{i\perp}^2}{2x_iM_0}. \notag \\
 \end{eqnarray}

In what follows, we adopt the notation $x=x_2$ and $x_1=1-x$.

\subsection{Form factors}

The hadron matrix element for the spin-1/2 to spin-1/2 processes can be expressed as:
\begin{eqnarray}
	&  & \langle \Sigma^0(P^{\prime}, \frac{1}{2},S_{z}^{\prime})|\bar{u}\gamma^{\mu}(1-\gamma_{5})c|\Xi_c^0(P, \frac{1}{2},S_{z})\rangle\nonumber \\
	& = & \int\{d^{3}p_{2}\}\frac{\phi^{\prime}(x^{\prime},k_{\perp}^{\prime})\phi(x,k_{\perp})}{2\sqrt{p_{1}^{+}p_{1}^{\prime+}(p_{1}\cdot\bar{P}+m_{1}M_{0})(p_{1}^{\prime}\cdot\bar{P}^{\prime}+m_{1}^\prime M_{0}^{\prime})}}\nonumber \\
	&  & \times\sum_{\lambda_{2}}\bar{u}(\bar{P}^{\prime},S_{z}^{\prime})\left[\bar{\Gamma}^{\prime}(\slashed p_{1}^{\prime}+m_{1}^{\prime})\gamma^{\mu}(1-\gamma_{5})(\slashed p_{1}+m_{1})\Gamma\right] \nonumber\\
	&& \times u(\bar{P},S_{z}).\label{eq:matrix_element_onehalf}
\end{eqnarray}

where
\begin{eqnarray}
&&m_{1}=m_{c},\quad m_{1}^{\prime}=m_{u},\quad m_{2}=m_{(di)},\quad\bar{P}=p_{1}+p_{2}, \nonumber\\
&& \bar{P}^\prime=p_{1}^{\prime}+p_{2},\quad M_{0}^2=\bar{P}^{2},\quad M_{0}^{\prime2}=\bar{P}^{\prime2}.\label{parameter}
\end{eqnarray}
The momenta of the $c$ quark in the initial baryon and $u$ quark in the final baryon are denoted by $p_1$ and $p_1^{\prime}$, respectively. The variable $p_2$ represents the momentum of the spectator diquark. The quantities $P$ and $P^{\prime}$ denote the four-momenta of the initial and final baryon states, respectively, while $M$ and $M^{\prime}$ correspond to their physical masses.

It is noted that  one can also define $M_{0}$ and $M^{\prime}_{0}$ for the initial and final baryon states, respectively, such that $\bar{P}^{(\prime)2}=M_0^{(\prime)2}$. The quantity $\bar\Gamma$ appearing in Equation~\eqref{eq:matrix_element_onehalf} is defined as:
\begin{eqnarray}
&&	\Gamma_{S}=\bar{\Gamma}^{\prime}_{S}=1,\notag\\
&&	\bar{\Gamma}^{\prime}_{A}  =\frac{1}{\sqrt{3}}\bigg(-\slashed\epsilon(p_{2},\lambda_{2}) \nonumber\\
&& \;\;\;\;\;\;\;\;\;\;\;\;\;\;\;+\frac{M_0^{\prime}+m_1^{\prime}+m_2}{\bar{P}^{\prime}\cdot p_2+m_2M_0^{\prime}}\epsilon(p_{2},\lambda_{2})\cdot\bar{P}^{\prime}\bigg)\gamma_{5}.\label{axial-vector diquarkprime}
\end{eqnarray}
By comparing the LFQM results for the hadron matrix element with Equation~\eqref{ff1/2}, we can extract the form factors by solving the following equations:
 \begin{widetext}
\begin{eqnarray}
&&	{\rm Tr}\{(\slashed{P'}+M')\bigg[\frac{M\gamma_\mu}{\bar M} f^{\frac{1}{2}\to\frac{1}{2}}_1(q^2)+\frac{P_\mu}{\bar M}f^{\frac{1}{2}\to\frac{1}{2}}_2(q^2)+\frac{P^\prime_\mu}{\bar M}f^{\frac{1}{2}\to\frac{1}{2}}_3(q^2)\bigg](\slashed{P}+M)\{\Gamma_{i}\}_\mu\}=H^{\frac{1}{2}}_i,\notag\\
&&	{\rm Tr}\{(\slashed{P'}+M')\bigg[\frac{M\gamma_\mu}{\bar M} g^{\frac{1}{2}\to\frac{1}{2}}_1(q^2)+\frac{P_\mu}{\bar M}g^{\frac{1}{2}\to\frac{1}{2}}_2(q^2)+\frac{P^\prime_\mu}{\bar M}g^{\frac{1}{2}\to\frac{1}{2}}_3(q^2)\bigg]\gamma_5(\slashed{P}+M)\{\Gamma_{5i}\}_\mu\}=K^{\frac{1}{2}}_i,\;i=1,2,3
\end{eqnarray}
where  $H^{\frac{1}{2}}_i$, $K^{\frac{1}{2}}_i$ are defined as
\begin{eqnarray} 
	H^{\frac{1}{2}}_i&=&\int\frac{dxd^2k_\perp}{2(2\pi)^3}\frac{\phi^\prime(x^\prime,k^\prime_\perp)\phi(x,k_\perp)}{2\sqrt{x^\prime_1x_1(p^\prime_1\cdot\bar{P}^\prime+m^\prime_1M^\prime_0)(p_1\cdot\bar{P}+m_1M_0)}}\nonumber\\
	&&\times{\rm Tr}\{(\slashed{\bar{P}}^\prime+M^\prime_0)\Gamma^\prime_{S(A)}(\slashed p^\prime_1+m^\prime_1)\gamma_\mu(\slashed p_1+m_1)\Gamma_{S(A)}(\slashed{\bar{P}}+M_0)\{\Gamma_i\}_\mu\}\notag\\
	K^{\frac{1}{2}}_i&=&\int\frac{dxd^2k_\perp}{2(2\pi)^3}\frac{\phi^\prime(x^\prime,k^\prime_\perp)\phi(x,k_\perp)}{2\sqrt{x^\prime_1x_1(p^\prime_1\cdot\bar{P}^\prime+m^\prime_1M^\prime_0)(p_1\cdot\bar{P}+m_1M_0)}}\nonumber\\
	&&\times{\rm Tr}\{(\slashed{\bar{P}}^\prime+M^\prime_0)\Gamma^\prime_{S(A)}(\slashed p^\prime_1+m^\prime_1)\gamma_\mu\gamma_5(\slashed p_1+m_1)\Gamma_{S(A)}(\slashed{\bar{P}}+M_0)\{\Gamma_{5i}\}_\mu\},
 \end{eqnarray}
 \end{widetext}
 
The different Dirac structures $\Gamma_i$ and $\Gamma_{5i}$ are
\begin{eqnarray} 
	&&\{\Gamma_{i}\}_\mu=\{\gamma_\mu,P_{\mu},P^\prime_{\mu}\},\nonumber\\
	&& \{\Gamma_{5i}\}_\mu=\{\gamma_\mu\gamma_5,P_{\mu}\gamma_5,P^\prime_{\mu}\gamma_5\},
\end{eqnarray} 
Then the form factors are solved as
 \begin{eqnarray}
 	&&f_1^{\frac{1}{2}\rightarrow\frac{1}{2}}=-\frac{\bar M(s_+H^{\frac{1}{2}}_1-2M^\prime H^{\frac{1}{2}}_2-2M H^{\frac{1}{2}}_3)}{4Ms_-s_+},
	\nonumber\\
	&& f_2^{\frac{1}{2}\rightarrow\frac{1}{2}}=\frac{\bar M(M^\prime s_+H^{\frac{1}{2}}_1-6M^{\prime 2}H^{\frac{1}{2}}_2+2(s_-+MM^\prime)H^{\frac{1}{2}}_3)}{2s_-s_+^2},\notag\\
	&&f_3^{\frac{1}{2}\rightarrow\frac{1}{2}}=\frac{\bar M(Ms_+H^{\frac{1}{2}}_1+2(s_-+MM^\prime)H^{\frac{1}{2}}_2-6M^2H^{\frac{1}{2}}_3)}{2s_-s_+^2}.  \nonumber\\\label{fffu}
\end{eqnarray}
Here, $s_\pm=(M\pm M^\prime)^2-q^2$, and the form factors $g^{\frac{1}{2}\to\frac{1}{2}}_i$ can be obtained by applying the following transformation:
\begin{eqnarray}
	&&M^{\prime}\to -M^{\prime}, \bar M\to \bar M,  H_{j}\to K_{j},   i=1\notag\\
	&&M^{\prime}\to -M^{\prime} , \bar M\to \bar M,  H_{j}\to -K_{j},  i=2,3.
\end{eqnarray} 

As the diquark can be either scalar or axial-vector, the physical baryon state is a combination of the baryon state with a scalar diquark and a state with an axial-vector diquark in LFQM. Therefore, the physical form factor can also be expressed as follows:
 \begin{eqnarray}
&& \rm  [formfactor]^{physical}=c_S \times  [formfactor]_S \nonumber\\
&& \;\;\;\;\;\;  +c_A\times  \rm  [formfactor]_A.
 \end{eqnarray}
 The coefficients $c_S$ and $c_A$ are flavor-spin factors that can be determined by the flavor-spin wave function of baryons in the diquark basis. The flavor-spin wave functions for spin-1/2 baryons are given by:
  \begin{eqnarray}
&&\Sigma^0=\frac{\sqrt{3}}{2}u[ds]_S -\frac{1}{2}u[ds]_A \notag\\
&&\Lambda=\frac{1}{2}u[ds]_S +\frac{\sqrt{3}}{2}u[ds]_A\notag\\
&&\Xi_c^0=c[d s]_S .
 \end{eqnarray}
 
The flavor wave functions of the diquark basis are given by:
\begin{eqnarray}
&&[q_1q_2]_A=\frac{1}{\sqrt{2}}(q_1q_2+q_2q_1),\nonumber\\
&& [q_1q_2]_S=\frac{1}{\sqrt{2}}(q_1q_2-q_2q_1).
\end{eqnarray}
therefore, the overlapping factors can be calculated and the results are presented in Table~\ref{ol}.

 %%%%%%%%%%%%%%%%%%%%%%%%%%%%%%%%%%%%
 \begin{table}[!htb]
\caption{The spin  factors for the baryon $\Xi^0_c$ decay induced by $c\to u$.}
\label{ol} %
\begin{tabular}{|c|c|c|c|c|c|c|c|c|}
\hline \hline
channel & $\quad c_S\quad$  & $\quad c_A\quad$   \tabularnewline
\hline
$\Xi_c^0\to\Sigma^0$&$\frac{\sqrt{3}}{2}$&$0$\tabularnewline
\hline
$\Xi_c^0\to\Lambda$&$\frac{1}{2}$&$0$\tabularnewline
\hline
\end{tabular}
\end{table}
%%%%%%%%%%%%%%%%%%%%%%%%%%%%%%%%%%%
 
\section{Numerical results}

After deriving the analytic expression in  the LFQM, we present the corresponding numerical results. For the calculation, we use the quark masses from Refs~\cite{Li:2010bb,Verma:2011yw,Shi:2016gqt}
\begin{eqnarray}
	 && m_u=m_d= 0.25~{\rm GeV}, m_s=0.37~{\rm GeV}, \nonumber \\
      &&m_c=1.4~{\rm GeV}.
\label{eq:mass_quark}
\end{eqnarray}
The masses of the diquarks can be approximated as
\begin{eqnarray}
	 m_{[ds]}=m_d+m_s.\label{eq:mass_diquark}
\end{eqnarray}
The masses of the baryons, their lifetimes, and the input parameter $\beta_{[ds]q}$ are presented in Table~\ref{mass_beta}.

%%%%%%%%%%%%%%%%%%%%%%%%%%%%%%%%%
\begin{table}[!htb]
\caption{baryon masses~\cite{Ghalenovi:2022dok,Workman:2022ynf} and other input parameters~\cite{Cheng:2018mwu,Kiselev:2001fw}.} 
\label{mass_beta} %
\begin{tabular}{|c|c|c|c|c|c|}
\hline \hline
Baryon & $\Xi_{c}^0$ & $\Sigma^0$&$\Lambda$ & $\eta$ & $\eta^{\prime}$
\tabularnewline
\hline
Mass $({\rm GeV})$&2.470&1.193& 1.116&0.548&0.958\tabularnewline
\hline
Lifetime$(\rm fs)$&151.9&&&&
\tabularnewline
\hline
\multicolumn{2}{|c|}{$\beta_{[ds]c}=0.58$}&\multicolumn{4}{|c|}{$ \beta_{[ds]u}=0.41$}\\\hline
\end{tabular}
\end{table}
%%%%%%%%%%%%%%%%%%%%%%%%%%%%%%%%%

\subsection{Form factors}

To analyze the $q^2$ dependence of the form factors in Eq.~\eqref{ff1/2}, we adopt the double-pole model parametrization scheme as follows:
 \begin{eqnarray}
	F(q^2)=\frac{F(0)}{1-\frac{q^2}{m_{\rm fit}^2}+\delta{(\frac{q^2}{m^2_{\rm fit}})}^2},
 \label{fit}
\end{eqnarray}
where $F(0)$ is the numerical results of form factor at $q^2=0$. And we takes $\{-0.000, -0.001, -0.005, -0.007, -0.01, -0.015\}$ for $q^2$ and fit the two parameters $m_{\rm fit}$ and $\delta$. Table~\ref{ffnr} presents the numerical results for the form factors and fitting parameters obtained using the double-pole model which parametrizes the $q^2$ dependence of the form factors.
Additionally, for  extrapolating the form factors to the full $q^2$ region,  we use the Bourrely-Caprini-Lellouch (BCL) parametrization~\cite{Boyd:1997kz,Caprini:1997mu,Bourrely:2008za,Bharucha:2010im} in which the form factors are expanded in powers of a conformal mapping variable. The BCL parametrization is shown as
 \begin{eqnarray}
f(q^2)&=&\frac{1}{1-q^2/m_R^2}\sum^{k_{max}}_{k=0} \alpha_kz^k(q^2,t_0),\nonumber\\
z(q^2,t_0)&=&\frac{\sqrt{t_+-q^2}-\sqrt{t_+-t_0}}{\sqrt{t_+-q^2}+\sqrt{t_+-t_0}},\notag\\ t_0&=&t_+\big(1-\sqrt{1-\frac{t_-}{t_+}}\big),\nonumber\\ 
t_\pm&=&(m_{{\cal B}_b}\pm m_{{\cal B}_c})^2.
\label{BCL}
\end{eqnarray}
The $m_R$ are the masses of the low-laying $D$ resonance.   

\begin{widetext}

 \begin{table}[!http]
\caption{Numerical results for the form factors of spin-1/2 to spin-1/2 $\Xi^0_c\to \Sigma(\Lambda)$ transitions. The F$(0)$ indicates the form factors when $q^2=0$. Results for the parameters $\delta$ and $m_{\rm fit}$ are obtained by fitting the form factors with the double-pole model  as in Eq.\eqref{fit}. The $\alpha_0$ and $\alpha_1$ are the parameters in the BCL model in Eq.~\eqref{BCL}. For the form factors $F(0)$ with $q^2=0$, we estimated their uncertainties caused by the parameters in LFQM, namely, $\beta_{c[ds]}$, $\beta_{u[ds]}$, and $m_{di}$, which are varied by $10\%$.}
 \label{ffnr} %
\begin{tabular}{|c|c|c|c|c|c|c|c|c|}
\hline \hline
\multirow{2}{*}{channel} &\multirow{2}{*}{form factor} & \multicolumn{3}{c|}{Pole model}&\multicolumn{2}{c|}{BCL model}
\cr\cline{3-7}
&&F(0) &$m_{\rm fit}$ & $\delta$  &$\alpha_0$ &$\alpha_1$\\
\hline
\multirow{6}{*}{$\Xi_{c}^0\to \Sigma^{0}$}&$ f_1$&$0.5988\pm 0.0765\pm 0.0500\pm 0.0407$&$1.98$&$0.40$&$0.5827$ & $0.9969$\cr\cline{2-7}
&$ f_2$&$0.1288\pm 0.0222\pm 0.0477\pm 0.0100$&$0.78$ &$0.50$ &$0.2784$ & $-9.2418$
\cr\cline{2-7}
&$ f_3$&$-0.5504\pm 0.1055\pm 0.0934\pm 0.0513$&$1.05$&$0.23$& $-0.8453$ & $18.2072$\cr\cline{2-7}
&$ g_1$&$0.3838\pm 0.0133\pm 0.0384\pm 0.0168$&$2.46$&$1.15$&$0.3432$& $2.5073$\cr\cline{2-7}
&$ g_2$&$-0.4279\pm 0.1518\pm 0.1303\pm0.0593$&$2.64$&$20.5$&$-0.3765$&$-3.1717$
\cr\cline{2-7}
&$ g_3$&$1.1520 \pm 0.1626\pm 0.2791\pm 0.1190$&$1.30$&$0.60$&$1.4584$&$-18.9181$\tabularnewline
\hline
\multirow{6}{*}{$\Xi_{c}^0\to \Lambda$}&$ f_1$&$0.3382\pm0.0432\pm0.0290\pm0.0229$&$2.12$&$0.70$&$0.3167$ & $1.1188$\cr\cline{2-7}
&$ f_2$&$0.1188\pm0.0170\pm0.0318\pm0.0081$&$0.90$ &$0.42$ &$0.2283$ & $-5.6985$
\cr\cline{2-7}
&$ f_3$&$-0.3471\pm0.0640\pm0.0583\pm0.0314$&$1.06$&$0.26$& $-0.5498$ & $10.5412$\cr\cline{2-7}
&$ g_1$&$0.2536\pm0.0093\pm0.0245\pm0.0108$&$2.32$&$0.64$&$0.2284$& $1.3106$\cr\cline{2-7}
&$ g_2$&$-0.2977\pm0.0751\pm0.0694\pm0.0319$&$1.66$&$1.48$&$-0.3204$&$1.1828$
\cr\cline{2-7}
&$ g_3$&$0.8241\pm0.0758\pm0.1591\pm0.0685$&$1.21$&$0.37$&$1.1416$&$-16.5172$\tabularnewline
\hline
\end{tabular}
\end{table}
\end{widetext}
\begin{widetext}

\begin{figure}[htbp!]
	\begin{minipage}[t]{0.4\linewidth}
		\centering
		\includegraphics[width=1\columnwidth]{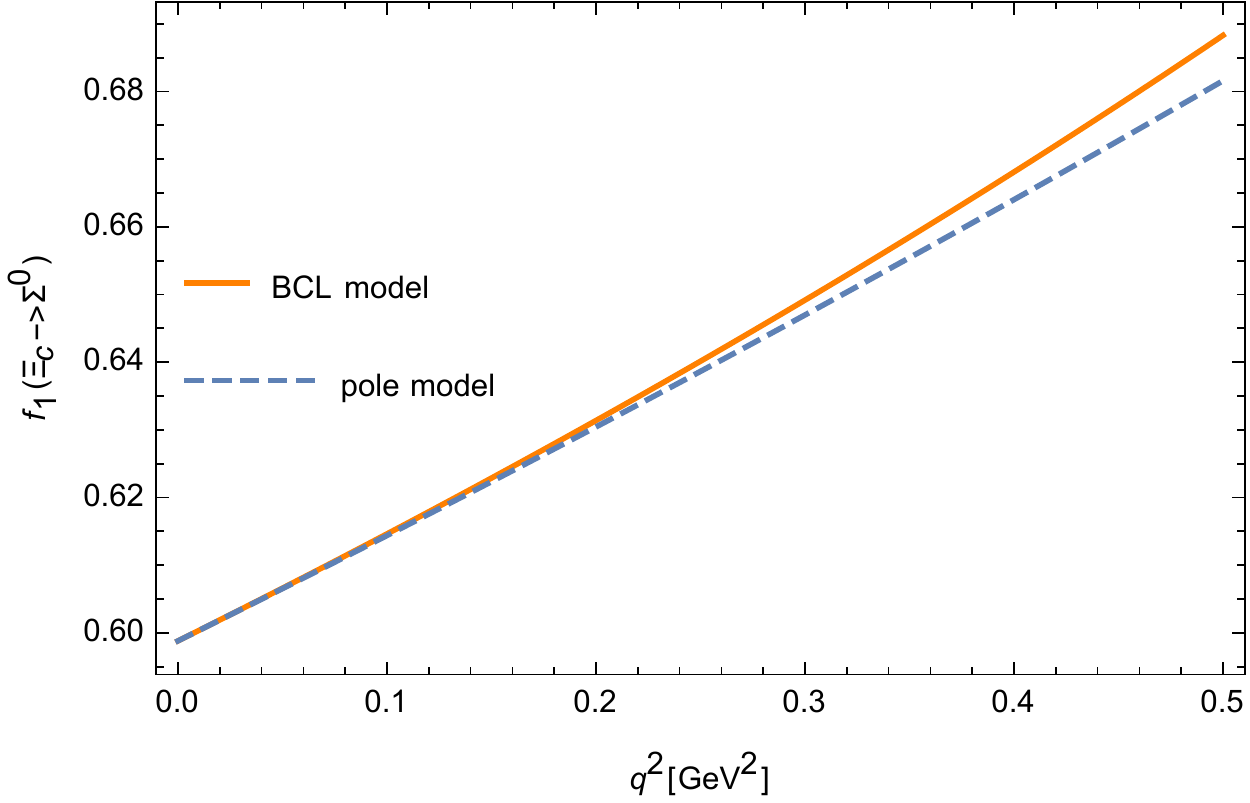}
	\end{minipage}
	\begin{minipage}[t]{0.4\linewidth}
		\centering		\includegraphics[width=1\columnwidth]{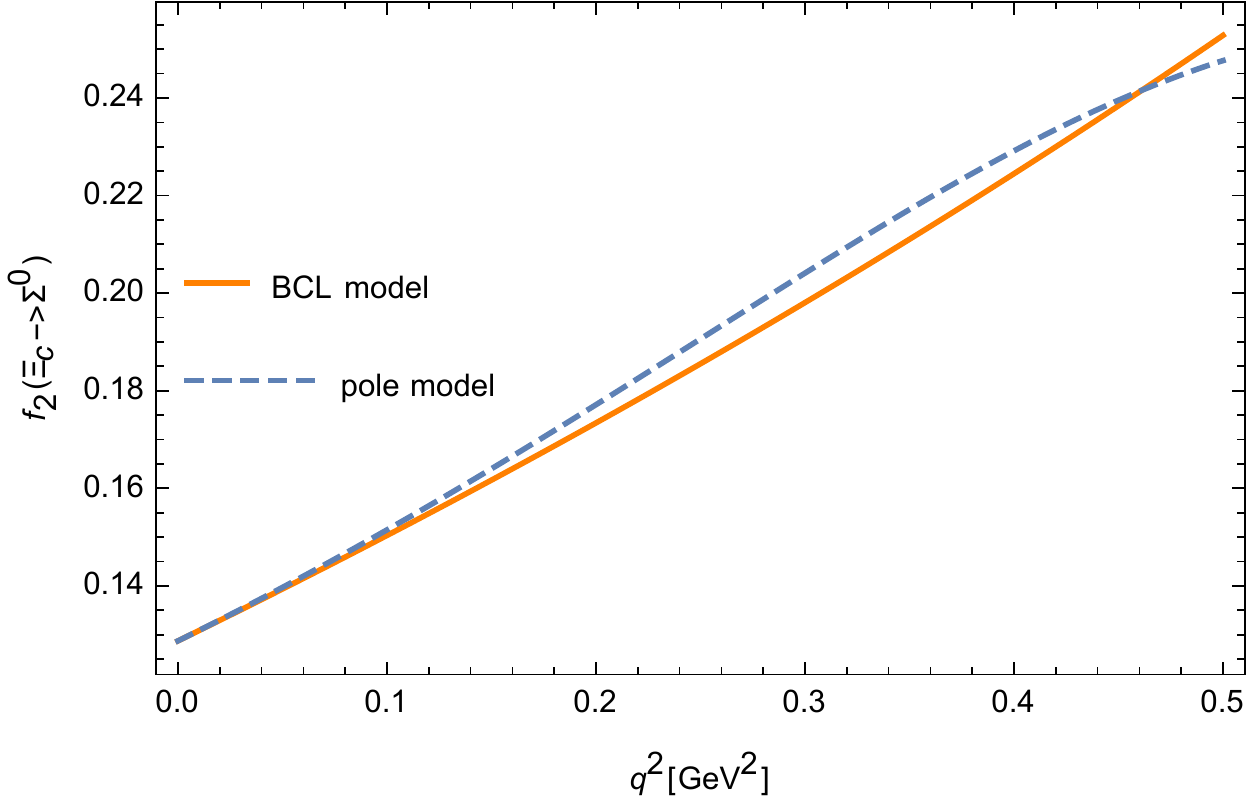}
	\end{minipage}
	\begin{minipage}[t]{0.4\linewidth}
		\centering
		\includegraphics[width=1\columnwidth]{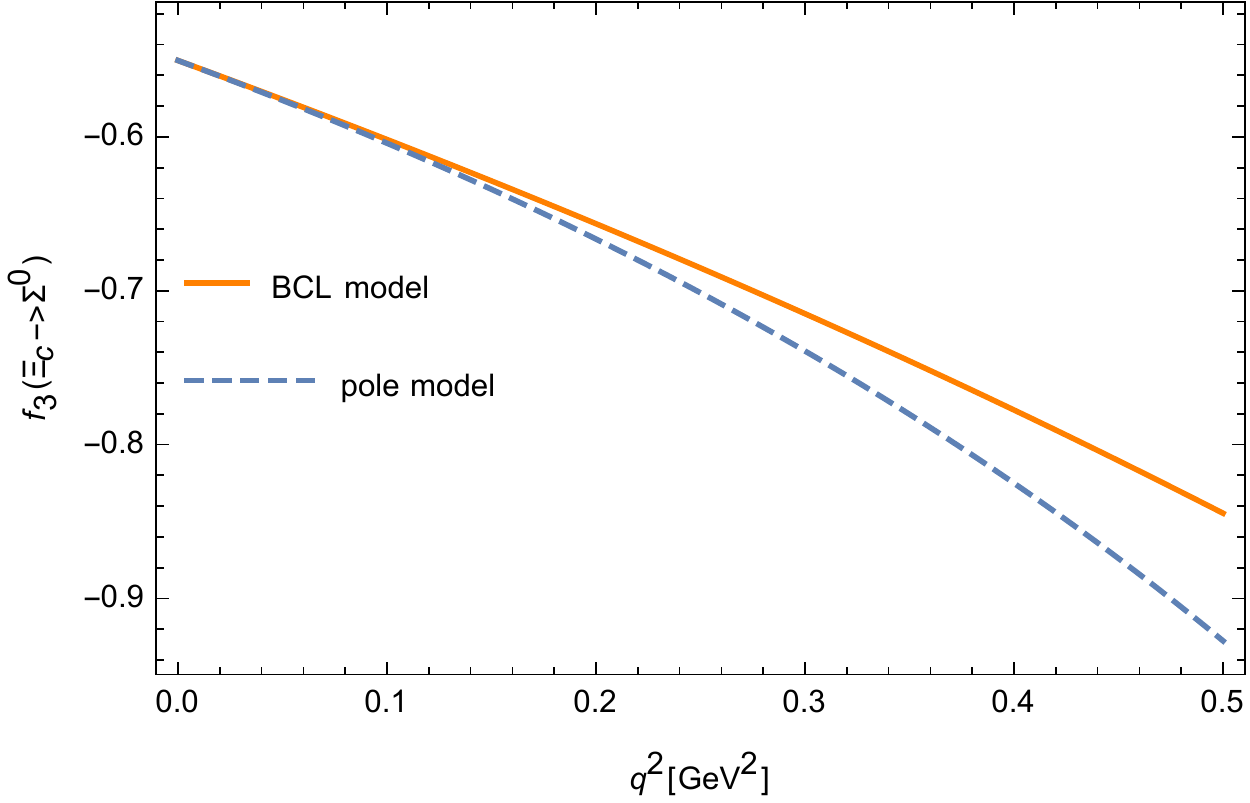}
	\end{minipage}
	\begin{minipage}[t]{0.4\linewidth}
		\centering
		\includegraphics[width=1\columnwidth]{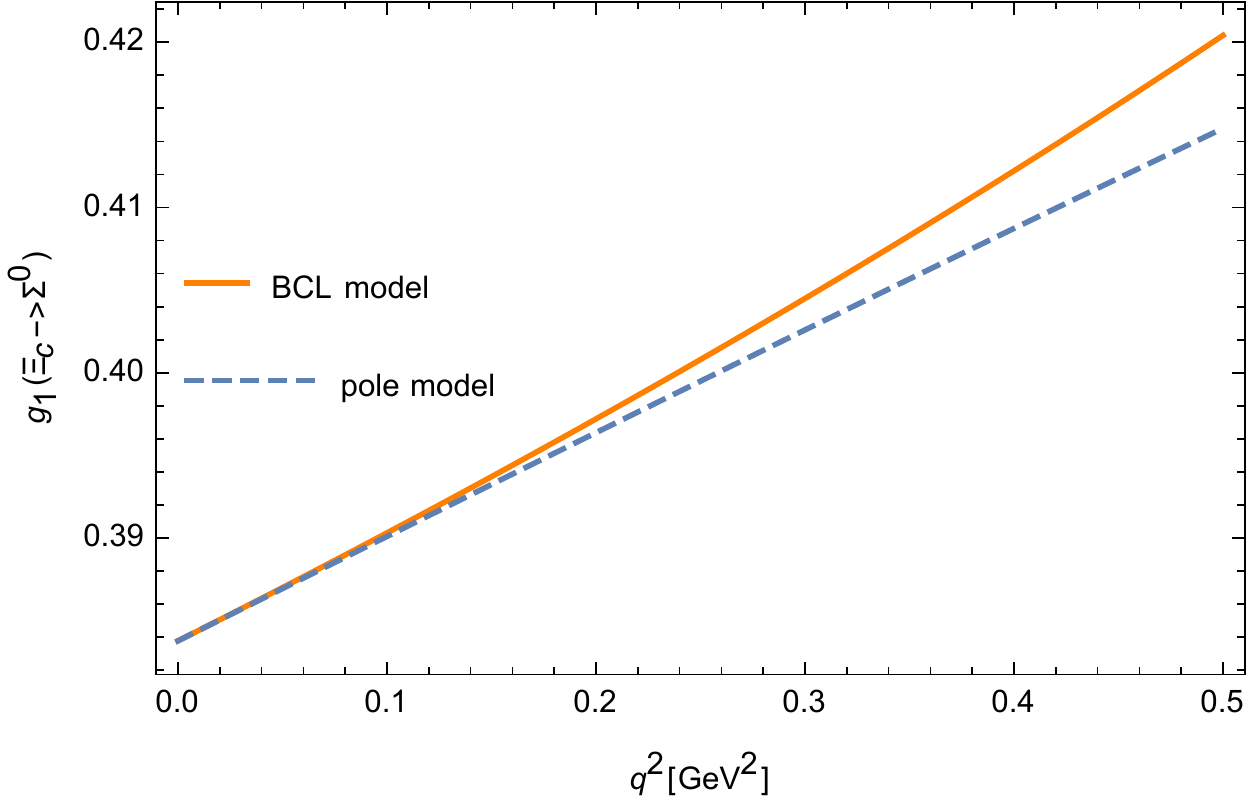}
	\end{minipage}
        \begin{minipage}[t]{0.42\linewidth}
		\centering
		\includegraphics[width=1\columnwidth]{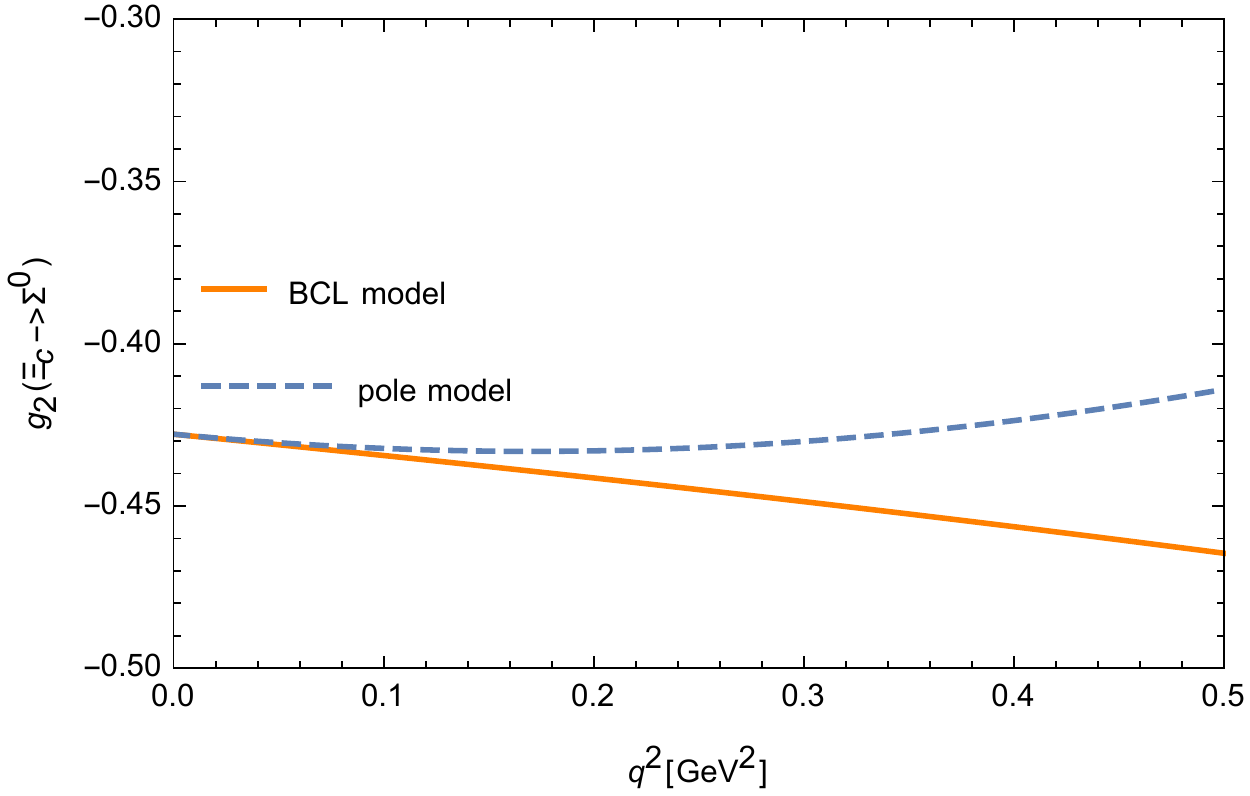}
	\end{minipage}
        \begin{minipage}[t]{0.40\linewidth}
		\centering
		\includegraphics[width=1\columnwidth]{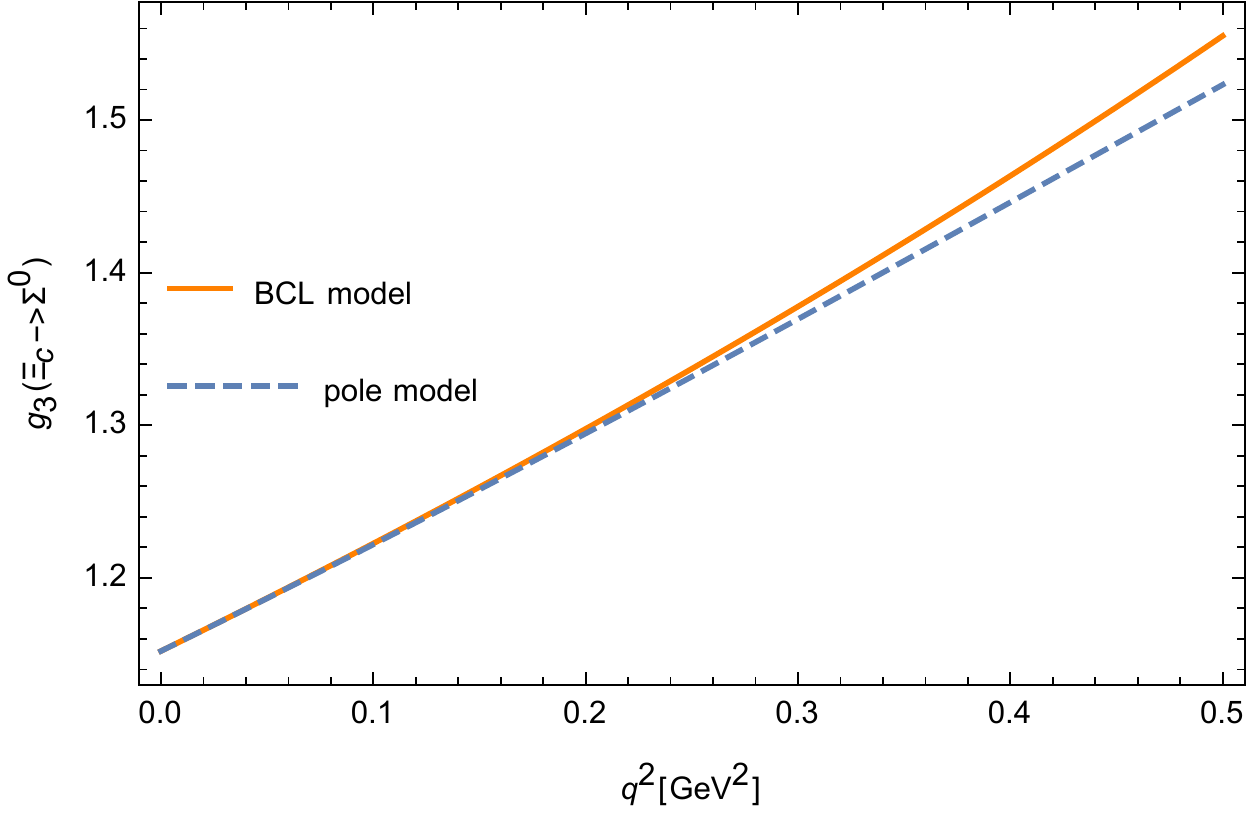}
	\end{minipage}
	\caption{ The $q^2$ dependence form factors of $\Xi^0_c\to\Sigma^0$ process with BCL model (Orange line) and pole model (blue line).}
\label{fig:1}
\end{figure} 
%%%%%%%%%%%%%%%%%%%%%%%%%%%%%%%%%%%%%%

\end{widetext}
To analyze the $q^2$ dependence of the form factors, we also plot the results of the form factors as functions of $q^2$ in Fig.~\ref{fig:1}. From Fig.~\ref{fig:1}, one can see that the fit results with two different models are broadly consistent with each other. However, the $q^2$-dependent form factor $g_2$ exhibits a large discrepancy with $q^2\sim (m_{{\cal B}_c}-m_{{\cal B}_u})^2$ for the two models. In the analysis presented in Ref.~\cite{Liu:2022mxv}, the form factor has a pole structure corresponding to the specific current. In our work, the pole mass $m_{pole}$ should be set to $m_{pole}=m_{D}$, which is consistent with the BCL model in Eq.\eqref{BCL}. However, the fit result of $g_2$ with the pole model is some different from our conclusion, especially for the $\Xi^0_c\to\Sigma^0$ process. Therefore,  it is likely that the BCL model  describes the $q^2$ dependence of the form factor better.
\subsection{Nonleptonic decays}
\begin{figure}[htbp!] 
\includegraphics[width=1.0\columnwidth]{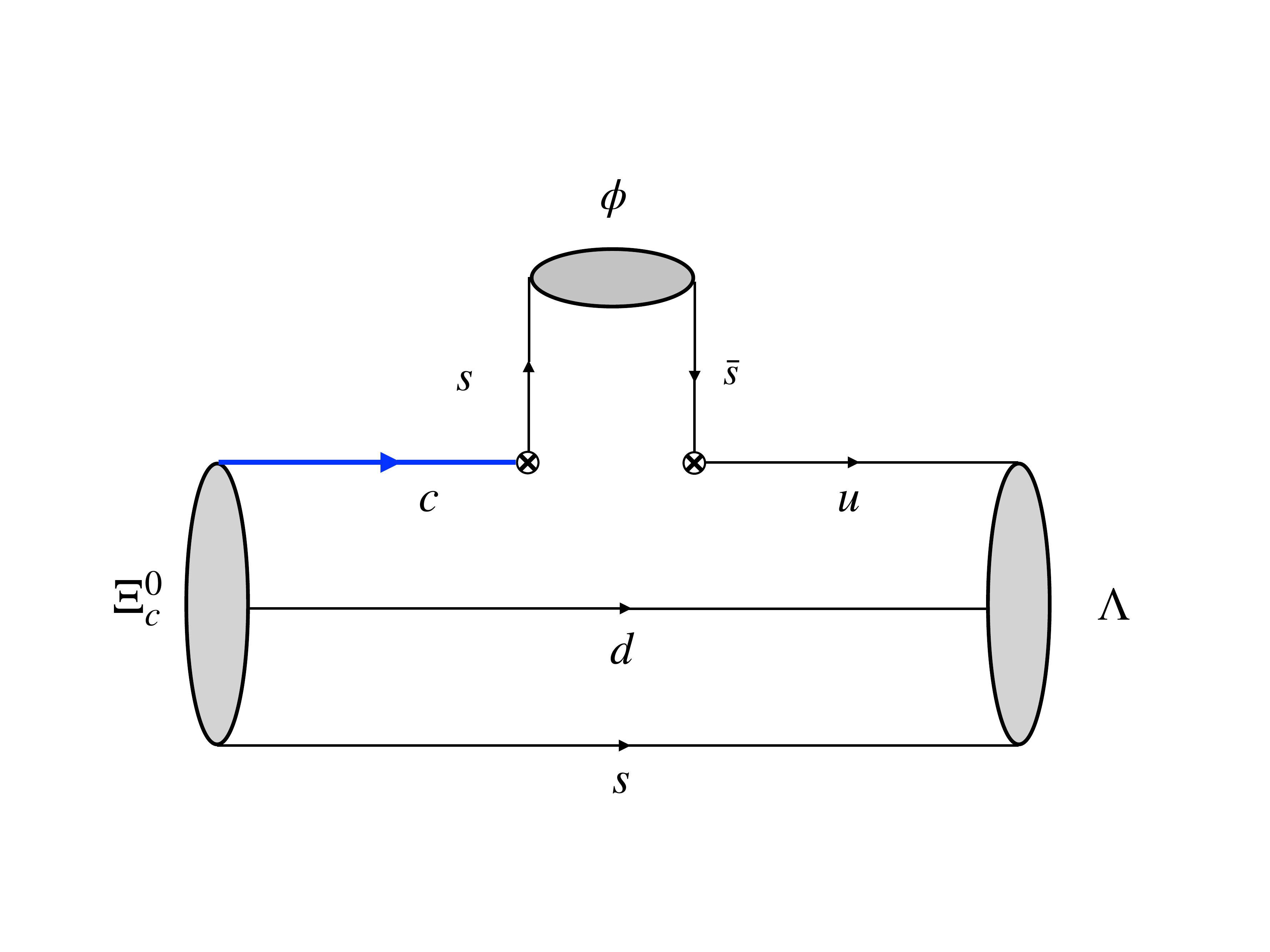}  
\caption{The Feynman diagrams for nonleptonic decays of the  decay  $\Xi^0_c \to \Lambda \phi$ we investigate.}
\label{fd2}
\end{figure}

%%%%%%%%%%%%%%%%%%%%%%%
Expressing the physical states $\eta$ and $\eta^{\prime}$ in the quark flavor basis, we have:
\begin{align}
    \left(
    \begin{array}{c}
       \left|\eta\right\rangle \\ \left|\eta'\right\rangle 
    \end{array}
    \right) = 
    \left(
    \begin{array}{cc}
       \cos\phi & -\sin\phi \\ \sin\phi & \cos\phi
    \end{array}
    \right)
    \left(
    \begin{array}{c}
       |\eta_q\rangle \\
       |\eta_s\rangle 
    \end{array}
    \right)
\end{align}
where $\phi$ denotes the mixing angle. Nonleptonic decay with $\eta^{(\prime)}$ emission can be estimated  with the helicity amplitude method. In Eq.~\eqref{amp}, the local matrix element  $\langle  \eta^{(\prime)}(P)|\bar s\gamma^{\mu}(1-\gamma_5)s| 0\rangle$ and  $\langle \eta^{(\prime)}(P)|\bar d\gamma^{\mu}(1-\gamma_5)d| 0\rangle$ can be expressed by the decay constant $f^{s}_{\eta^{(\prime)}}$ and $f^{d}_{\eta^{(\prime)}}$ as~\cite{Cheng:1998uy,Ali:1997nh,Yu:2022ngu}
\begin{eqnarray}
&&\langle 0|\bar q\gamma^{\mu}(1-\gamma_5)q| \eta_q(P)\rangle=\frac{i}{\sqrt{2}}f_qP^{\mu},\notag\\
&& \langle 0|\bar s\gamma^{\mu}(1-\gamma_5)s| \eta_s(P)\rangle=i f_sP^{\mu}
\end{eqnarray}
where all the other parameters are listed below
\begin{eqnarray}
&&m_{\eta}=0.548~{\rm GeV}, m_{\eta^{\prime}}=0.958~{\rm GeV},\notag\\
&&f_q=1.07f_\pi, f_s=1.34f_\pi,\notag\\
&&f_{\pi}=0.130~{\rm GeV}, \phi=39.3^{\circ},
\end{eqnarray}
Then the amplitude of nonleptonic decays becomes
\begin{widetext}
\begin{eqnarray}
&&i\mathcal{M}(\Xi_c^0\to \Sigma^0\eta)=\frac{iG_F}{\sqrt{2}}P_{\eta}^{\mu}(\frac{\cos{\phi}}{\sqrt{2}}V_{cd}V^*_{ud}f_q-\sin{\phi}V_{cs}V^*_{us}f_s)\notag\\
&&\;\;\;\;\;\;a_2\langle \Sigma^0(P^\prime,S_z^\prime)|\bar u \gamma_\mu(1-\gamma_5)c|\Xi^0_c(P,S_z)\rangle\notag\\
&&i\mathcal{M}(\Xi_c^0\to \Sigma^0\eta^{\prime})=\frac{iG_F}{\sqrt{2}}P_{\eta^{\prime}}^{\mu}(\frac{\sin{\phi}}{\sqrt{2}}V_{cd}V^*_{ud}f_q+\cos{\phi}V_{cs}V^*_{us}f_s)\notag\\
&&\;\;\;\;\;\;a_2\langle \Sigma^0(P^\prime,S_z^\prime)|\bar u \gamma_\mu(1-\gamma_5)c|\Xi^0_c(P,S_z)\rangle
\end{eqnarray}
\end{widetext}
The helicity amplitudes in the nonleptonic decay processes are defined as
\begin{eqnarray}
&&HV^{S}_{\lambda,\lambda^\prime}=\langle \Sigma^0(P^\prime,\lambda^\prime)|\bar u \slashed{P}_{\eta^{(\prime)}} c|\Xi^0_c(P,\lambda)\rangle,\notag\\
&& HA^{S}_{\lambda,\lambda^\prime}=\langle \Sigma^0(P^\prime,\lambda^\prime)|\bar u \slashed{P}_{\eta^{(\prime)}} \gamma_5 c|\Xi^0_c(P,\lambda)\rangle.
\end{eqnarray}
With the help of helicity amplitude, the total decay width of the spin-1/2 to spin-1/2 processes can be expressed as
\begin{eqnarray}
	&&\Gamma^{\frac{1}{2}\to\frac{1}{2}}={\cal P_{\eta^{(\prime)}}}\frac{\sqrt{s_+s_-}}{32\pi M^3}\bigg(|H_{\frac{1}{2},-\frac{1}{2}}^{\frac{1}{2}}|^2+|H_{-\frac{1}{2},\frac{1}{2}}^{\frac{1}{2}}|^2\bigg),\notag\\
	&&{\cal P_{\eta}}=\frac{G_F^2}{2}((\frac{\cos{\phi}}{\sqrt{2}}V_{cd}V^*_{ud}f_q)^2+(\sin{\phi}V_{cs}V^*_{us}f_s)^2) a_2^2,\notag\\
    && {\cal P_{\eta'}}=\frac{G_F^2}{2}((\frac{\sin{\phi}}{\sqrt{2}}V_{cd}V^*_{ud}f_q)^2+(\cos{\phi}V_{cs}V^*_{us}f_s)^2) a_2^2.\notag\\
\end{eqnarray}
where $H^{S}_{\lambda,\lambda^\prime}=HV^{S}_{\lambda,\lambda^\prime}-HA^{S}_{\lambda,\lambda^\prime}$. The expressions of the nonleptonic helicity amplitudes are shown in Appendix.~\ref{A}. 
Although we haven't accounted for the nonfactorizable contributions, we can estimate their impact by varying the value of $N_c$. A method outlined in Ref. [47] suggests that adjusting the value of $N_c$ in the Wilson coefficient $a_2 = C_1 + C_2/N_c$ can help estimate these contributions in B decays. Since there are no experimental results for these two specific processes at present, we can use a similar process, namely $\Xi^0_c \to \Lambda \phi$~\cite{Cheng:2018hwl}, which does have experimental data available, to determine the value of $N_c$. To determine the appropriate value for $N_c$, we examine the corresponding Feynman diagram, shown in Fig.~\ref{fd2}.

So the way we compute the branching ratio for this process. we vary the value of $N_c$ and determine that the branching ratio matches the experimental result of $4.9\pm 1.5 \times 10^{-4}$. Specifically, we find that the branching ratio is equal to the experimental value when $N_c=5.4\pm 1.2$.
Using the formulas above and choosing $N_c=5.4\pm 1.2$,  we give numerical results of hadron nonleptonic two-body decays in
Table~\ref{fq}. 
\begin{widetext}

\begin{table}[!htb]
\caption{Numerical results of decay width and branching fraction in heavy baryon nonleptonic decays using BCL model $f(q^2)$ at $N_c=5.4\pm 1.2$. We have assessed the uncertainties arising from variations in $N_c$ and the parameters within LFQM, namely, $\beta_{c[ds]}$, $\beta_{u[ds]}$, and $m_{di}$, each of which was subject to a $10\%$ variation.}\label{fq} 
\begin{tabular}{|c|c|c|c|c|c|c|c|c|}
\hline \hline
channel & $\Gamma(\times10^{-16}\rm{GeV})$   & Br($10^{-4}$)   \tabularnewline
\hline
$\Xi^0_{c}\to \Sigma^{0}\eta$&$4.85\pm1.62\pm0.51\pm0.39\pm0.34$&$1.12\pm0.38\pm0.12\pm0.09\pm0.08$\tabularnewline
\hline
$\Xi^0_{c}\to \Sigma^{0}\eta^{\prime}$&$6.64\pm2.22\pm0.92\pm0.76\pm0.85$&$1.53\pm0.51\pm0.21\pm0.18\pm0.19$\tabularnewline
\hline
\end{tabular}
\end{table}

Using the formulas above,  we give numerical results of the non-leptonic two-body decay processes $\Xi^0_{c}\to \Sigma^{0}\eta^{(')}$  in Table~\ref{fq1} and $\Xi^0_{c}\to \Lambda \eta^{(')}$ in Table~\ref{fq}.

\begin{table}[!htb]
\caption{Numerical results of decay width and branching fraction in heavy baryon nonleptonic decays using BCL model $f(q^2)$ at $N_c=5.4\pm1.2$. We have assessed the uncertainties arising from variations in $N_c$ and the parameters within LFQM, namely, $\beta_{c[ds]}$, $\beta_{u[ds]}$, and $m_{di}$, each of which was subject to a $10\%$ variation.}\label{fq1} %
\begin{tabular}{|c|c|c|c|c|c|c|c|c|}
\hline \hline
channel & $\Gamma(\times10^{-16}\rm{GeV})$   & Br($10^{-4}$)   \tabularnewline
\hline
$\Xi^0_{c}\to \Lambda\eta$&$1.70\pm0.57\pm0.18\pm0.14\pm0.13$&$0.39\pm0.13\pm0.04\pm0.03\pm0.03$\tabularnewline
\hline
$\Xi^0_{c}\to \Lambda\eta^{\prime}$&$2.53\pm0.85\pm0.35\pm0.27\pm0.31$&$0.58\pm0.19\pm0.08\pm0.06\pm0.07$\tabularnewline
\hline
\end{tabular}
\end{table}

\end{widetext}

We can roughly estimate the non-factorization effect by varying the $N_c$.  This strategy is used in Ref.~\cite{Ali:1998eb}, it has been suggested that the  non-factorizable contributions in $B$ decays can be estimated through a variation of $N_c$ in Wilson coefficient $a_1$ or $a_2$. The effects are estimated in Table~\ref{nonlepton}. The results show that for this kind of processes $\Xi^0_{c}\to \Sigma^{0}\eta/\eta'$, the decay width and branching fraction depend significantly on $N_c$. To explore less model-dependent observables, we can utilize the ratios of branching ratios for certain decay channels as a means to study the variations in decay widths resulting from different values of $N_c$. Therefore one can define the value ${\mathcal R}_{\Sigma^0(\Lambda)}$ as
\begin{eqnarray}
&&{\mathcal R}_{\Sigma^0}=\frac{\Gamma(\Xi^0_c\to \Sigma^0 \eta)}{\Gamma(\Xi^0_c\to \Sigma^0 \eta')},\notag\\
&&{\mathcal R}_{\Lambda}=\frac{\Gamma(\Xi^0_c\to \Lambda \eta)}{\Gamma(\Xi^0_c\to \Lambda \eta')}.
\end{eqnarray}
The values of these ratios are shown in Table~\ref{nonlepton}. We observed that this ratio exhibits a high degree of independence from the parameter $N_c$, indicating reduced model dependence in the observables. This enhanced predictability makes it a valuable candidate for experimental testing.

\begin{widetext}
\begin{center}
\begin{table}[!htb]
\caption{Numerical results of decay width and branching fraction in doubly heavy baryon nonleptonic decays using $f(q^2)$ when $N_c=2,3,5.4,\infty$.}\label{nonlepton} %
\begin{tabular}{|c|c|c|c|c|c|c|c|c|c|c|c|c|}
\hline \hline
&\multicolumn{3}{|c|}{$N_c=2$}&\multicolumn{3}{|c|}{$N_c=3$}&\multicolumn{3}{|c|}{$N_c=5.4$}&\multicolumn{3}{|c|}{$N_c=\infty$}\tabularnewline
\hline
channel & $\Gamma(\times10^{-18}\rm{GeV})$   & Br($10^{-6}$)&$\mathcal{R}$& $\Gamma(\times10^{-16}\rm{GeV})$   & Br($10^{-4}$)&$\mathcal{R}$ & $\Gamma(\times10^{-16}\rm{GeV})$   & Br($10^{-4}$)&$\mathcal{R}$ &$\Gamma(\times10^{-16}\rm{GeV})$& Br($10^{-4}$) &$\mathcal{R}$\tabularnewline
\hline
$\Xi^0_{c}\to \Sigma^{0}\eta$&$4.44$&$1.02$&$\multirow{2}{*}{0.73}$&$1.14$&$0.26$&$\multirow{2}{*}{0.72}$&$4.85$&$1.12$&$\multirow{2}{*}{0.73}$&$13.12$&$3.02$&$\multirow{2}{*}{0.73}$\tabularnewline

$\Xi^0_{c}\to \Sigma^{0}\eta'$&$6.08$&$1.40$&&$1.56$&$0.36$&&$6.64$&$1.53$&&$17.96$&$4.13$&\tabularnewline
\hline
$\Xi^0_{c}\to \Lambda\eta$&$1.56$&$0.36$&$\multirow{2}{*}{0.68}$&$0.40$&$0.09$&$\multirow{2}{*}{0.64}$&$1.70$&$0.39$&$\multirow{2}{*}{0.67}$&$4.61$&$1.06$&$\multirow{2}{*}{0.68}$\tabularnewline

$\Xi^0_{c}\to \Lambda\eta'$&$2.31$&$0.53$&&$0.59$&$0.14$&&$2.53$&$0.58$&&$6.83$&$1.57$&\tabularnewline
\hline
\end{tabular}
\end{table}
\end{center}   
\end{widetext}
\section{Summary}
%%%%%%%%%%%%%%%%%%%%%%%
The two-body hadronic decays decays of the baryons $\Xi^{0}_c \to \Sigma^0\eta^{(')}$ and $\Xi^{0}_c \to \Lambda\eta^{(')}$ are studied in this work. In this work, we employed the light-front-quark model to study nonleptonic decays of the baryons $\Xi^0_c$. Specifically, we utilized the diquark picture, where the two spectator heavy quarks can be approximated as a scalar or an axial-vector diquark, and treated the baryon state as a meson state. We obtained the form factors using the hadron matrix element and used them to estimate the decay widths and branching fractions of two-body nonleptonic decays. \\
We have used the helicity amplitudes to obtain the phenomenological results, which include the predicted branching fractions. With the lifetime of 
 $\Xi^0_c$  presented in Table~\ref{mass_beta}, we have calculated the branching fractions and listed them in Table~\ref{fq}. Lastly, we explore the dependence on $N_c$ and introduce new observations denoted as $\mathcal{R}$, which exhibit minimal sensitivity to parameters. These quantities are less model-dependent, rendering them suitable for experimental testing. This study represents an exploratory endeavor, taking into account the non-perturbative nature of matrix elements. To delve deeper into these non-perturbative matrix elements, we also plan to employ the first-principle lattice method for conducting additional calculations. 
The obtained phenomenological results are helpful in the search for such types of decay processes in future experiments. 
%%%%%%%%%%%%%%%%%%%%%%%
\section*{Acknowledgements}
%%%%%%%%%%%%%%%%%%%%%%%
We thank Zhi-Peng Xing, Prof. Wei Wang, Prof. Xiao-Gang He, Meng-Lin Du, C. W. Liu, and Zhen-Xing Zhao for their useful discussions. 
This work was supported in part by NSFC under Grant Nos.12147147, 11735010, 11911530088, U2032102, 12125503.

\appendix
\section{Helicity amplitude}\label{A}
The helicity amplitudes for the spin-1/2 to spin-1/2 nonleptonic decay processes can be expressed in terms of the form factors defined in the hadron matrix element. Here, we define $(M^2-M^{\prime2})\pm q^2=\hat{M}_q^{\pm}$. The helicity amplitudes can then be written as follows:
\begin{eqnarray}
&&HV_{-\frac{1}{2},\frac{1}{2}}^{\frac{1}{2}}=\frac{-i\sqrt{s_+}m_{\eta}}{2\bar{M}\sqrt{q^2}}[2M\bar{M}f^{\frac{1}{2}\to\frac{1}{2}}_1+\hat{M}_q^{+}f^{\frac{1}{2}\to\frac{1}{2}}_2\notag\\
&&+\hat{M}_q^{-}f^{\frac{1}{2}\to\frac{1}{2}}_3],\notag\\
&&HV_{-\lambda,-\lambda^\prime}^{\frac{1}{2}}=HV_{\lambda,\lambda^\prime}^{\frac{1}{2}}\label{A1},\\
% \end{eqnarray}
%  and 
% \begin{eqnarray}
&&HA_{-\frac{1}{2},\frac{1}{2}}^{\frac{1}{2}}=\frac{-i\sqrt{s_-}m_{\eta}}{2\bar{M}\sqrt{q^2}}[2M(M+M^\prime)g^{\frac{1}{2}\to\frac{1}{2}}_1\notag\\
&&\qquad\qquad\quad-\hat{M}_q^{+}g^{\frac{1}{2}\to\frac{1}{2}}_2-\hat{M}_q^{-}g^{\frac{1}{2}\to\frac{1}{2}}_3],\notag\\
&&HA_{-\lambda,-\lambda^\prime}^{\frac{1}{2}}=-HA_{\lambda,\lambda^\prime}^{\frac{1}{2}}\label{A2}\;.
\end{eqnarray}

\end{document}